\documentclass[aps,pra,twocolumn,superscriptaddress,showpacs,floatfix]{revtex4}
\usepackage{graphicx,dcolumn,bm,hyperref,amsmath,amssymb,xspace,epsfig,float,array,multirow,amsfonts,color}
\usepackage[normalem]{ulem}

\DeclareMathAlphabet{\mathpzc}{OT1}{pzc}{m}{it}

\begin{document}

\title{Quantum boomerang effect: beyond the standard Anderson model}

\author{L. Tessieri}
\email{luca.tessieri@umich.mx}
\affiliation{Instituto de F\'{\i}sica y Matem\'{a}ticas, Universidad Michoacana
  de San Nicol\'{a}s de Hidalgo, 58060, Morelia, Mexico}
\author{Z. Akdeniz}
\email{zehra.akdeniz@pirireis.edu.tr}
\affiliation{Faculty of Science and Letters, P\^{\i}r\^{\i} Reis University,
34940 Tuzla, Istanbul, Turkey}
\author{N. Cherroret}
\email{nicolas.cherroret@lkb.upmc.fr}
\affiliation{Laboratoire Kastler Brossel, Sorbonne Universit\'e, CNRS, ENS-PSL Research University, Coll\`ege de France, 4 Place Jussieu, 75005 Paris, France}
\author{D. Delande}
\email{dominique.delande@lkb.upmc.fr}
\affiliation{Laboratoire Kastler Brossel, Sorbonne Universit\'e, CNRS, ENS-PSL Research University, Coll\`ege de France, 4 Place Jussieu, 75005 Paris, France}
\author{P. Vignolo}
\email{patrizia.vignolo@inphyni.cnrs.fr}
\affiliation{Universit\'e C\^ote d'Azur, CNRS, Institut de Physique de Nice,
1361 route des Lucioles
06560 Valbonne, France}

\date{23rd February 2021}

\begin{abstract}
It was recently shown that wavepackets with skewed momentum distribution
exhibit a boomerang-like dynamics in the Anderson model due to Anderson localization:
after an initial ballistic motion, they make a U-turn and eventually come back to
their starting point.
In this paper, we study the robustness of the quantum boomerang effect  in various kinds of
disordered and dynamical systems: tight-binding models with pseudo-random potentials,
systems with band random Hamiltonians, and the kicked rotor.
Our results show that the boomerang effect persists in models with
pseudo-random potentials. It is also present in the kicked rotor, although
in this case with a specific dependency on the initial state.
On the other hand, we find that random hopping processes inhibit any drift motion of
the wavepacket, and consequently the boomerang effect.
In particular, if the random nearest-neighbor hopping amplitudes have zero average,
the wavepacket remains in its initial position.
\end{abstract}

\maketitle

\section{Introduction}

Anderson localization plays a key role in the physics of
disordered systems and inhomogeneous materials. In general terms, any
wave propagating in a random medium experiences multiple scattering and
localization occurs as a consequence of the destructive interference 
between the scattered partial waves.
The interference mechanism underlying localization explains why the
phenomenon affects not only quantum particles~\cite{Ande58} but any
kind of wave propagating in a random medium~\cite{She06}, including
atomic~\cite{Roati2008,Billy2008}, acoustic~\cite{May01}, and
electromagnetic waves~\cite{Laurent2007,Kuhl08}.

The quantum ``boomerang effect'' constitutes a recent development in the field
of Anderson localization~\cite{Prat2019}. The authors of Ref.~\cite{Prat2019}
studied the dynamics of wavepackets in the  Anderson model.
They considered a wavepacket with momentum distribution peaked around a
non-zero mean momentum and  found that, after an initial ballistic drift,
the wavepacket moves back to its initial position and eventually gets
localized there.

The purpose of this paper is to verify whether the boomerang
effect is an exclusive feature of the Anderson model or, on
the contrary, exists also in related physical systems.
More specifically, we numerically investigate three classes
of models: 1)~Anderson-like models with pseudo-random potentials,
2) tight-binding models with random hopping amplitudes that connect
the first $b$ nearest neighbors, and 3) the quantum kicked rotor, a
paradigm of quantum chaos known to exhibit Anderson localization in
momentum space.
Models of the first class have the same Schr\"{o}dinger equation as
the Anderson model, but the site energies are pseudo-random, rather
than strictly random, variables. Hamiltonians of the second group
are described by band random matrices and constitute a
natural generalization of the tridiagonal Anderson model with
purely diagonal disorder. As for the kicked rotor, finally, it
can be formally mapped onto a tight-binding model with a band Hamiltonian
and pseudo-random site energies.
These three kinds of models are selected to shed light on the role
played by three specific features of the Anderson model, namely,
the truly random nature of the site energies, the deterministic
character of hopping amplitudes, and their short (actually, nearest-neighbor)
range.
Models of the first and the third class are investigated to demonstrate
that the quantum boomerang effect survives when the potential is pseudo-random.
The band random matrices and  kicked-rotor models also allow us to explore the
role of hopping processes having a random character or extending beyond nearest
neighbors. While the long-range but deterministic hopping terms in the kicked rotor
do not suppress the boomerang dynamics of the wavepacket, we find that random hopping
amplitudes in tight-binding models destroy it.

In detail, our first system is a tridiagonal ``Anderson model'' with pseudo-random
site energies, originally proposed in~\cite{Griniasty1988}.
By varying a single parameter of this model, one can change
the spatial correlations of the site energies and drastically
alter the transport properties of the system, which can cross over
from metal to insulator, with an intermediate regime
in which the system is not spatially homogeneous on average~\cite{Thouless1988,Fishman1992}.
Our numerical simulations show that the boomerang effect takes place in the
insulating regime and even persists in the intermediate regime, though its properties
are not universal and depend on specific parameters of the pseudo-random disorder.

To evaluate the effect of off-diagonal disorder, we further consider
band random matrices of the form proposed in~\cite{Fyodorov1991},
namely, matrices with zero-average random elements in a central band
made up of $2b+1$ diagonals and vanishing elements $H_{ij} = 0$
for $|i-j| > b$. For this class of matrices we find that the
boomerang effect disappears: the wavepacket spreads before eventually getting
localized, but its center of mass does not move.
We also consider a variant of the previous model, in which the random
elements of the first subdiagonals have a nonzero
average. This corresponds to a Hamiltonian including a nonzero Laplacian
term in addition to the band random matrix component. We find that the Laplacian term
is essential for the existence of the quantum boomerang effect, which survives as long as the
deterministic contribution to nearest-neighbor hopping dominates over the off-diagonal
random terms. When the width of the band or the random hopping amplitudes
are increased, the off-diagonal disorder takes over and the boomerang effect  vanishes. 

Our last benchmark system, the kicked rotor, can be mapped onto the Anderson
model with pseudo-random site energies and non-random but long-range
hopping, with the effective band width being determined by
the strength of the kick potential~\cite{Fishman1982,Casati1989}.
In such a model, localization occurs in momentum space, as was
confirmed by experiments with cold atoms~\cite{Raizen1994,Chabe2008}.
Our numerical analysis shows that the  the kicked rotor also exhibits the
quantum boomerang effect. Nevertheless, we find that the boomerang dynamics
significantly depends on the choice of the initial state, a phenomenon without
parallel in the Anderson model.

The paper is organized as follows. In Sec.~\ref{themodel} we summarize the main
results obtained in~\cite{Prat2019} for the boomerang effect in the standard one-dimensional Anderson model.
The Anderson model with pseudo-random site energies is then analyzed in
Sec.~\ref{pseudo}.
Sec. \ref{bands} is devoted to band random matrices, while we discuss the boomerang effect in the kicked-rotor model in Sec.~\ref{kicked}.
Sec.~\ref{concl} concludes the paper.

\section{The boomerang effect in the Anderson model}
\label{themodel}

The standard one-dimensional (1D) Anderson model is defined by the Hamiltonian
\begin{equation}
  H = \sum_{n=-\infty}^{\infty} \bigg [ -J(|n \rangle \langle n+1|
  + |n \rangle \langle n - 1|)
  + |n \rangle \varepsilon_{n} \langle n| \bigg ] .
  \label{H1}
\end{equation}
In Eq.~(\ref{H1}), $J$ is the hopping amplitude.
The site energies $\varepsilon_{n}$ are independent,
identically distributed random variables with box distribution
\begin{equation}
	p(\varepsilon) = \left\{ \begin{array}{cl}
		1/2W & \mbox{ for } -W \leq \varepsilon \leq W \\
		0 & \mbox{otherwise} . \\
	\end{array} \right.
	\label{box_dist}
\end{equation}
Note that the average value of the energies vanishes, $\overline{
\varepsilon_{n} }~=~0$, while the variance of the disorder
is
\begin{displaymath}
	\sigma_{\varepsilon}^{2} = \overline{\varepsilon_{n}^{2}} =
	\frac{W^{2}}{3} .
\end{displaymath}
In the previous expressions, as in the rest of this paper, we use a
vinculum to denote the ensemble average of a random variable,
i.e.,
\begin{displaymath}
\overline{x}= \int x\ p(x)\ \mathrm{d} x.
\end{displaymath}

In Ref.~\cite{Prat2019}, the authors considered the time evolution
in the Anderson model~(\ref{H1}) of a Gaussian wavepacket:
\begin{equation}
    \psi(x_{n},t=0) = {\cal N}
    \exp \left[- \frac{x_{n}^2}{2\sigma_{x}^{2}(0)} + i k_{0} x_{n} \right] ,
	\label{init_wp}
\end{equation}
where $x_{n} = n d$, $d$ is the lattice spacing, and ${\cal N}$ is a
normalization constant (with ${\cal N} \simeq d/\sqrt{\pi \sigma_{x}^{2}(0)}$
for $\sigma_x(0) \gg d$).
The wavepacket~(\ref{init_wp}) has a momentum distribution which is also a
Gaussian, centered around $k_{0}$ and of width $\sigma_{p}(0) \sim
\hbar/\sigma_{x}(0)$.
To guarantee that the dynamics in disorder is governed by a well-defined energy
$E \simeq - 2J \cos(k_{0}d)$, Prat and coworkers~\cite{Prat2019} considered
wavepackets with a narrow momentum distribution corresponding to relatively
large values of $\sigma_{x}(0)$.

In~\cite{Prat2019}, it was found that the quantum evolution of the 
wavepacket resembles that of a boomerang: after initially moving
away from the origin, the center of mass of the wavepacket performs
a U-turn and eventually returns to its initial position.
While its center of mass moves in this boomerang-like fashion,
the wavepacket spatially spreads in an asymmetric fashion, with the
symmetry being eventually restored at long times when the dynamics is
completely halted by Anderson localization.
In Ref.~\cite{Prat2019} it was also shown that the drift and spreading
of the wavepacket are connected through the dynamical relationship
\begin{equation}
  \frac{d}{dt} \langle x^{2}(t) \rangle = 2 v_{0} \langle x(t) \rangle
  \label{dd}
\end{equation}
where $v_0$ is the mean wavepacket velocity, given by 
\begin{equation}
	v_{0} = \frac{2Jd}{\hbar} \sin(k_0 d).
	\label{v0}
\end{equation}
In Eq.~(\ref{dd}), the symbols $\langle x(t) \rangle$ and $\langle x^{2}(t) \rangle$
stand for the first two moments of the disorder-averaged density distribution
$\overline{|\psi(x_{n},t)|^{2}}$, i.e.
\begin{equation}
	\begin{array}{ccl}
		\langle x(t) \rangle & = & \displaystyle
		\sum_{n} x_{n} \overline{|\psi(x_{n},t)|^{2}} ,\\
		\langle x^{2}(t) \rangle & = & \displaystyle
		\sum_{n} x_{n}^{2} \overline{|\psi(x_{n},t)|^{2}} =\sigma_{x}^{2}(t).
	\end{array}
	\label{two_moments}
\end{equation}

\begin{figure}[H]
\begin{center}
\includegraphics[width=0.8\linewidth]{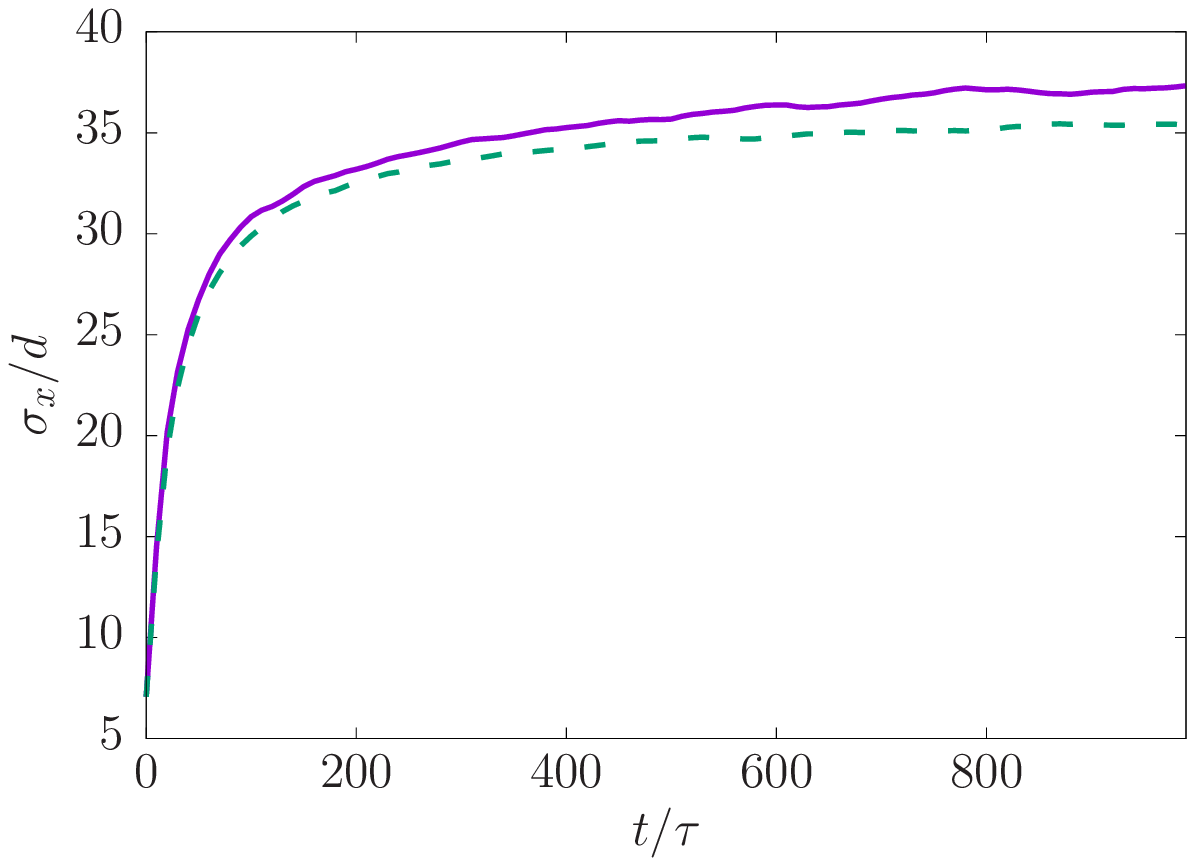}
\includegraphics[width=0.8\linewidth]{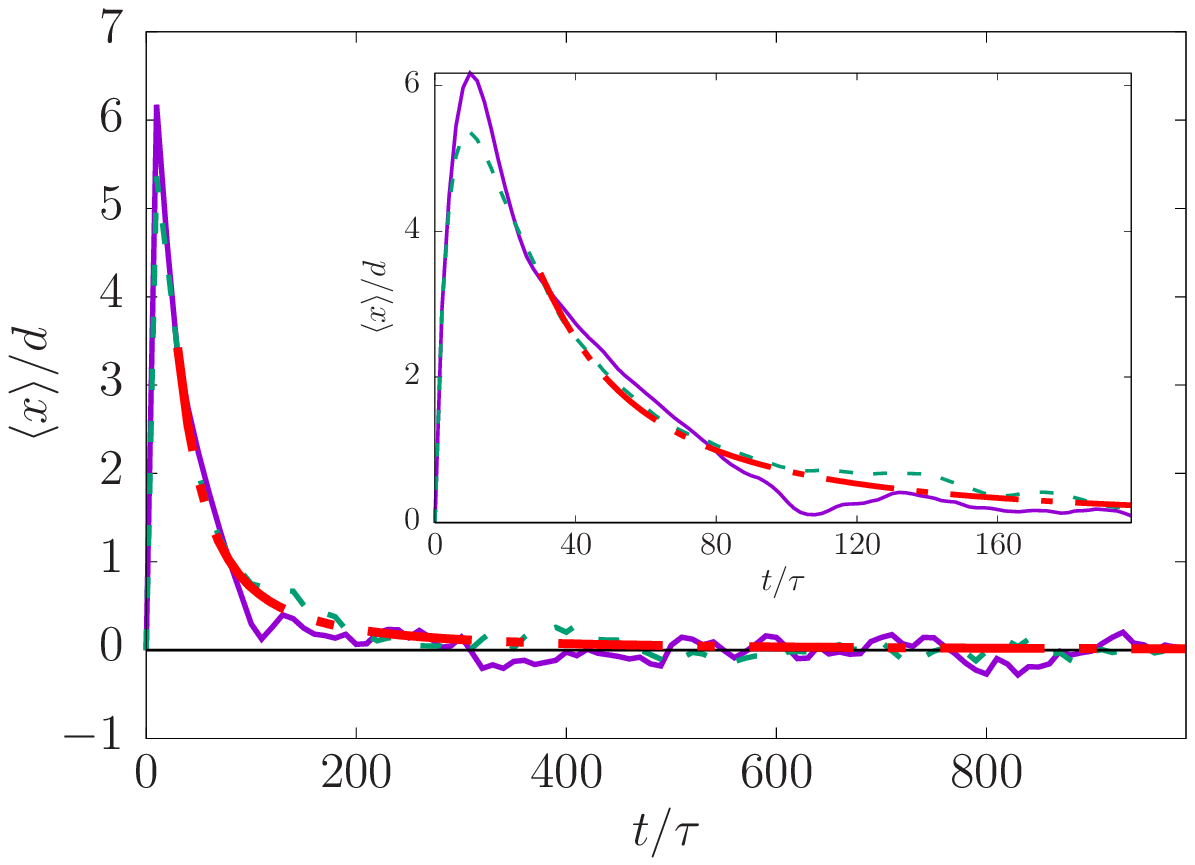}
\includegraphics[width=0.8\linewidth]{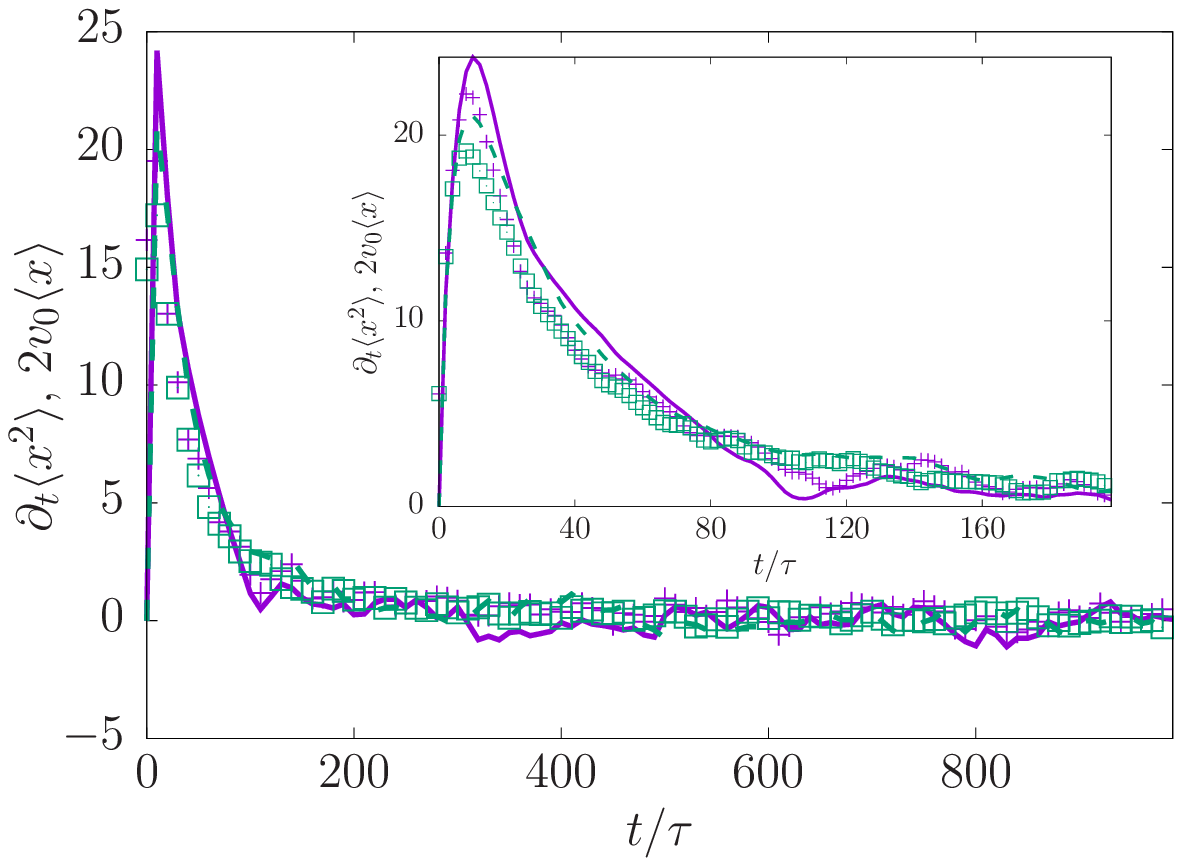}
\caption{\label{fig1}
  From top to bottom: wavepacket width $\sigma_x=\sqrt{\langle x^2\rangle}$,
  wavepacket center of mass $\langle x\rangle$,
  and comparison of both sides of Eq.~(\ref{dd}) as functions of the rescaled
  time $t/\tau$, with $\tau=\hbar/J$, for the standard Anderson model (green dashed lines) and its
  pseudo-random counterpart with $\gamma=3$ (violet continuous lines).
  The red dot-dashed curve in the middle panel corresponds to the asymptotic
  expression (\ref{Prat-DD}), while the horizontal black line marks the zero of the vertical axis.
  In the lower panel the symbols (squares for the Anderson model and crosses
  for the pseudo-random analogue) correspond to the term $d\langle x^{2} \rangle/ dt$,
  while the continuous lines represent the term $ 2 v_0 \langle x\rangle$ (in units of
  $d^2/\tau$).
  The data were obtained by averaging over $N_{c} = 2000$ disorder configurations.
  The error bars, not shown in the figure, have an amplitude of $\sim 1$ in each panel
  in the corresponding unit.} 
\end{center}
\end{figure}
The time evolution of the first two moments~(\ref{two_moments}) and of
both sides of Eq.~(\ref{dd}) (green dotted curves) is shown in Fig.~\ref{fig1}.
The numerical data in Fig.~\ref{fig1} were obtained 
for a disorder strength $\sigma_{\varepsilon}^{2} = J^2/3$ (corresponding
to $W=J$) and for a wavepacket of the form~(\ref{init_wp}) with
$k_0d=1.4$, and $\sigma_{x}(0) = 10d$. The ensemble averages were performed
over $N_{c}=2000$ disorder realizations.

To describe the temporal evolution of the center of mass, it is convenient to introduce the mean scattering time
\begin{displaymath}
	\tau_{\ell} = v_{0} \ell ,
\end{displaymath}
where $v_{0}$ is given by Eq.~(\ref{v0}) while $\ell$ is the mean free
path, which in the 1D model~(\ref{H1}) is equal to one fourth of the localization
length $\ell_{loc}$, i.e., $\ell = \ell_{loc}/4$.
The latter typically controls the asymptotic spatial decay of the envelope of the wavefunction
and is defined as 
\begin{displaymath}
	\ell_{loc} = \left[ \lim_{N \to \infty} \frac{1}{Nd} \sum_{n=1}^{N} \ln \left|
	\frac{\psi(x_{n+1})}{\psi(x_{n})} \right| \right]^{-1} .
\end{displaymath}
For weak disorder, the localization length can be computed
analytically~\cite{Luck1989,Izrailev1999}. For an eigenstate of energy $E = -2 J\cos (kd)$ one
has 
\begin{equation}
\ell_{loc}^{-1} = \frac{1}{d} \frac{\langle \varepsilon_{n}^{2} \rangle}{8J^2 \sin^{2} (kd)}
\left[ 1 + \sum_{l=1}^{\infty} \frac{\langle \varepsilon_{n} \varepsilon_{n+l} \rangle}
{\langle \varepsilon_{n}^{2} \rangle} \cos \left( 2 l k d \right) \right] .
\label{Born}
\end{equation}
Note that, when disorder is uncorrelated, the term in the square brackets in the
right-hand side (r.h.s.) of Eq.~(\ref{Born}) reduces to unity.
As long as the momentum distribution of the wavepacket is sufficiently narrow, only momenta close to $k\simeq k_0$ contribute to the dynamics, so that the time evolution of the quantum boomerang effect is essentially governed by the single time scale $\tau_\ell=\ell_{loc}(k\simeq k_0)/4v_0$. Under this condition, a simple analytical expression for
the center-of-mass position $\langle x(t) \rangle$ was derived in \cite{Prat2019} in the limit of long times $t\gg\tau_\ell$,
namely,
\begin{equation}
  \langle x(t) \rangle \simeq 64 \ell \left( \frac{\tau_{\ell}}{t} \right)^{2}
  \ln \frac{t}{4 \tau_{\ell}}.
  \label{Prat-DD}
\end{equation}
Eq.~(\ref{Prat-DD}) matches well the result obtained with numerical simulations,
as can be seen from the central panel of Fig.~\ref{fig1}, in which the analytical
expression~(\ref{Prat-DD}) is represented by the dot-dashed red line.

\section{Pseudo-random potentials}
\label{pseudo}

In this section, we analyze the boomerang effect in a variant of the
Anderson model~(\ref{H1}) in which the random site energies are
replaced by pseudo-random variables.
Models of this kind appear naturally in the study of dynamical systems
like the kicked rotor~\cite{Chi81} and for this reason were studied
in~\cite{Griniasty1988,Thouless1988,Fishman1992}.
Our purpose here is to establish whether the boomerang effect survives when the
site energies are pseudo-random variables given by
\begin{equation}
  \varepsilon_{n} = \mathcal{W} \cos \phi_{n},
  \label{pseudo_random}
\end{equation}
with
\begin{equation}
	\phi_{n} = \pi \sqrt{5} n^{\gamma} .
	\label{phase}
\end{equation}
Site energies of the form~(\ref{pseudo_random}) have vanishing average
$\overline{\varepsilon_{n}}~=~0$ and variance equal to
$\overline{ \varepsilon_{n}^2}=\mathcal{W}^{2}/2$. Here and in the
rest of this section, we use the symbol $\overline{(\cdots)}$ to denote the
average taken over a sequence of variables, i.e.,
\begin{displaymath}
\overline{x_{n}} = \lim_{N \to \infty} \frac{1}{N} \sum_{n=1}^{N} x_{n} .
\end{displaymath}
Extensive studies of the model~(\ref{H1}) with site energies~(\ref{pseudo_random})
have shown  that the extended or localized character of the eigenstates
depends crucially on the parameter $\gamma$ in
Eq.~(\ref{phase})~\cite{Griniasty1988,Thouless1988,Fishman1992}.
Specifically, all states are localized if $\gamma \ge 2$, while there are extended states
if $0 < \gamma \le 1$. 
For the intermediate range $1 < \gamma < 2$ the potential has a slowly varying
period for large values of the site index $n$. In this regime the state at the band center
is delocalized, while the other states are localized but with a longer localization length
than for the corresponding random model.

With the aim to compare the Anderson model with its pseudo-random analogue,
we set $\mathcal{W} = J\sqrt{2/3}$ in order to have the same disorder
strength for the two models.
In the weak-disorder limit this implies, in particular, the same value of
the mean free path.
For our numerical calculations, we considered finite chains of $N_{s} =
2000$ sites. For each chain, we let the initial wavepacket~(\ref{init_wp})
evolve in time. We finally averaged over $N_{c} = 2000$ different chains,
obtained with a shift of the site energies~(\ref{pseudo_random}).
More specifically, we took site energies for the $i$-th chain
of the form
\begin{displaymath}
	\varepsilon_{n}^{(i)} = \mathcal{W} \cos \phi_{n}^{(i)} 
\end{displaymath}
with
\begin{equation}
	\phi_{n}^{(i)} = \pi \sqrt{5} \left[ n + 10 (i - 1) \right]^{\gamma} .
	\label{shifted_phase}	
\end{equation}

We show in Fig.~\ref{fig1} the numerical results obtained with this model
for $\gamma=3$, (continuous violet curves). We observe that the width of the wavepacket
and its center of mass evolve in time exactly in the same way regardless of whether
the site energies are random or pseudo-random variables. This is fully consistent with
the conclusion reached in previous studies~\cite{Griniasty1988, Fishman1982} 
that for $\gamma \ge 2$ the eigenstates of the model~(\ref{H1})
localize in the same way when the truly random site energies are replaced
by the variables~(\ref{pseudo_random}).

\begin{figure}
\begin{center}
\includegraphics[width=0.8\linewidth]{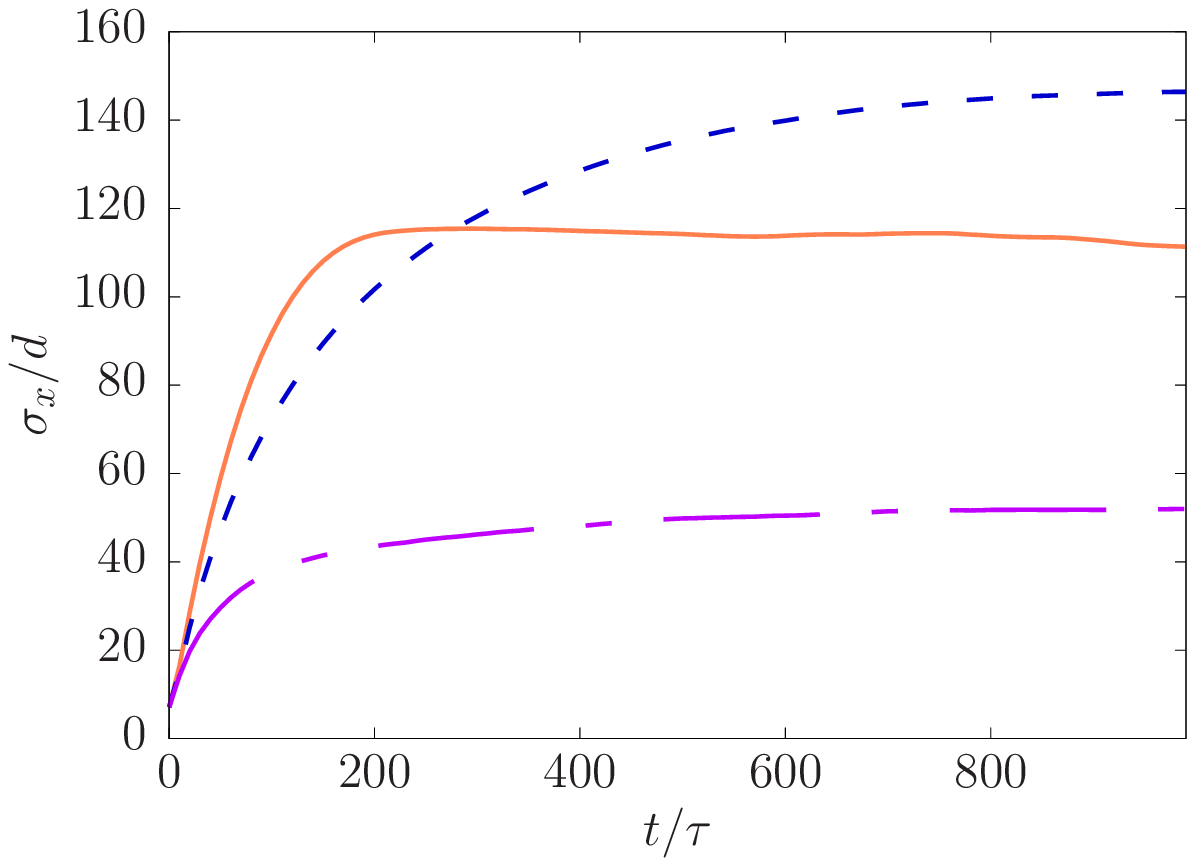}
\includegraphics[width=0.8\linewidth]{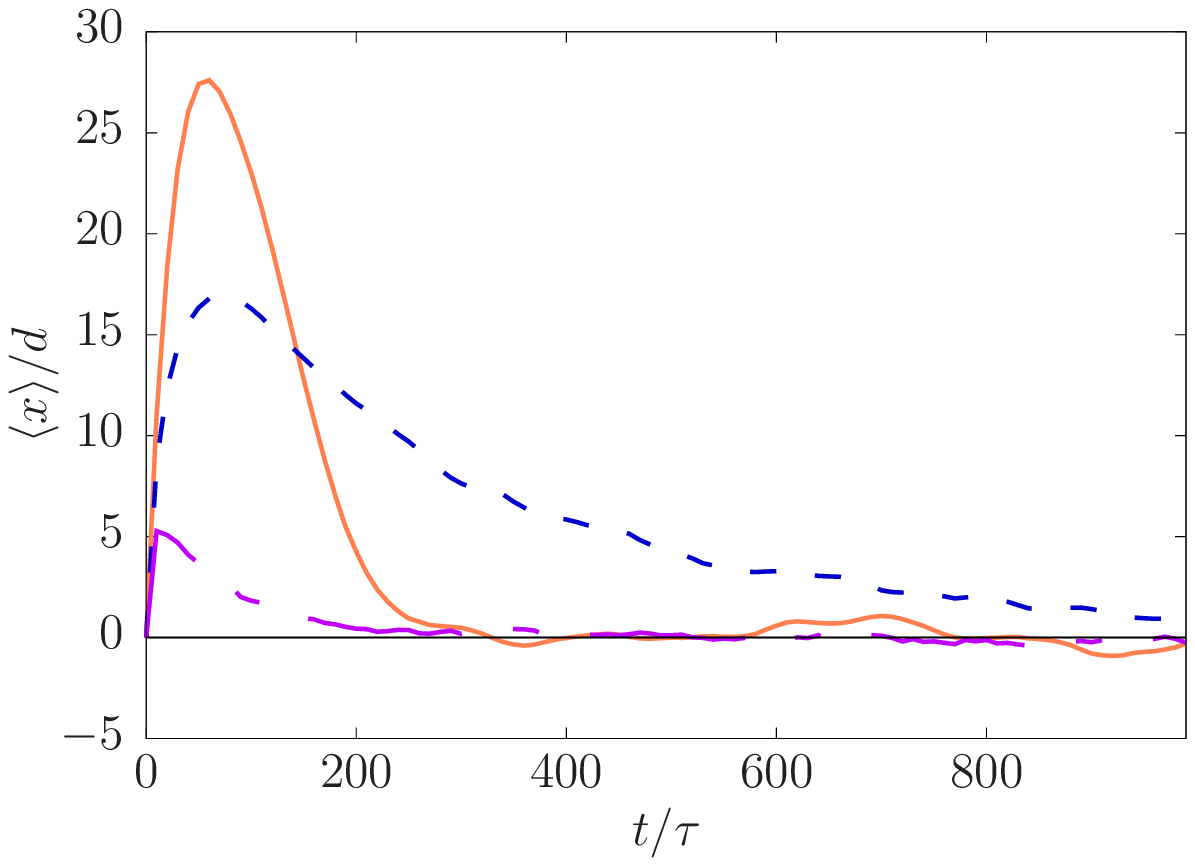}
\includegraphics[width=0.8\linewidth]{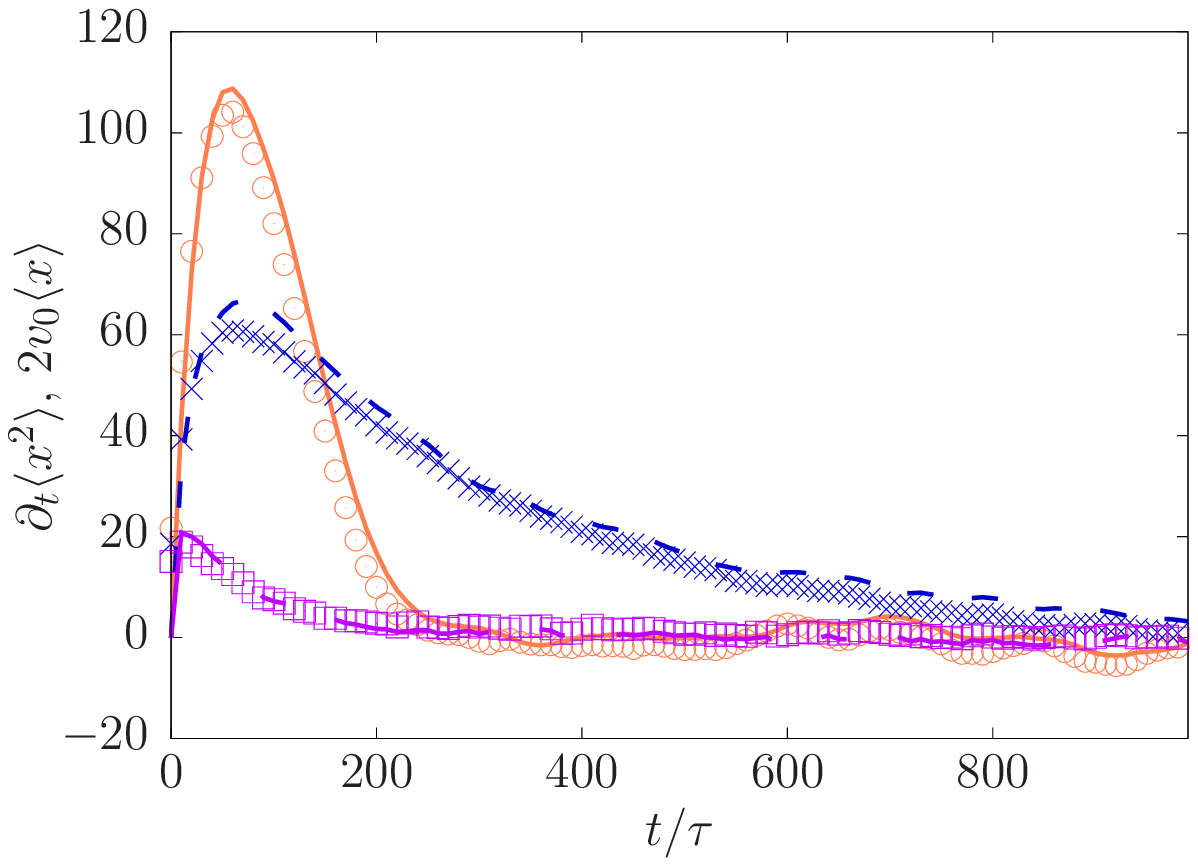}
\caption{\label{fig2}
  From top to bottom: wavepacket width $\sigma_x$, wavepacket center of mass
  $\langle x\rangle$,
  and comparison of both sides of~Eq. (\ref{dd}) as functions of the rescaled time
  $t/\tau$, with $\tau=\hbar/J$ for pseudo-random energies with $\gamma=1.4$
  and $i_{0}=0$ (light-salmon continuous line), $i_{0}=20000$ (blue dashed line), and
  $i_{0}=30000$ (violet dot-dashed line).
  In the bottom panel the symbols (circles for the case $i_{0}=0$, crosses
  for $i_{0}=20000$ and squares for $i_{0}=30000$)
  refer to $d\langle x^2\rangle/dt$, while the continuous lines
  represent $ 2 v_0 \langle x\rangle$.
  The data are averaged over $N_{c}=2000$ configurations. The error bars,
  not shown in the figure, have an amplitude of $\sim 1$ in each panel in the
  corresponding unit.}
\end{center}
\end{figure}

When $\gamma=1.4$, on the other hand, we are in the intermediate region $1<\gamma<2$
and the random lattice has long stretches of strongly correlated site energies
while the eigenstates are localized only over large spatial scales.
We find that, for $\gamma = 1.4$, the results can vary significantly from chain to chain,
depending on the value of the shift parameter $i$ in Eq.~(\ref{shifted_phase}).
This is illustrated by the plots in Fig.~\ref{fig2}, where we show
the data obtained for $\gamma=1.4$ by averaging over three groups of $N_{c} = 2000$
configurations. These configurations were obtained by letting the index $i$ vary in
the range $[i_{0},i_{0}+N_{c}]$ with $i_{0} = 0$ for the first group of
configurations, $i_{0} = 20000$ for the second one, and $i_{0} = 30000$ for the
last ensemble.
Comparing the results to those for a truly random lattice, we observe that the wavepacket
spreads more rapidly and its center of mass explores a larger part of the lattice
(compare with Fig.~\ref{fig1}).
Fig.~\ref{fig2} also emphasizes that spatial homogeneity on average is broken in the regime
$1 < \gamma < 2$: varying the shift $i_{0}$ changes the dynamics of both the variance and the
center of mass of the wavepacket. An increase of $i_{0}$, however, does not have an univocal
impact on the dynamics, as it can either enhance or reduce the delocalization
of the wavefunction.
Nevertheless, our results show that localization persists and the boomerang effect is
still present. Furthermore, as shown in the lower panel of Fig.~\ref{fig2}, we also find that
formula~(\ref{dd}) works within the numerical errors (not shown in
Fig.~\ref{fig2}).

\section{Band random matrices}
\label{bands}

Both the Anderson model considered in Sec.~\ref{themodel} and its pseudo-random
counterpart discussed in Sec.~\ref{pseudo} are tight-binding models with
nearest-neighbor bonds. In the study of quantum chaos and localization,
considerable attention has been given to a generalization of the Anderson model, in
which the Hamiltonian is a band random matrix (BRM) rather than a tridiagonal
matrix with purely diagonal disorder.
BRMs were originally introduced by Wigner~\cite{Wig55, Wig57}, but their
application in problems of quantum chaos and localization began in the late '80s and
early '90s~\cite{Sel85,Cas90a,Cas90b,Cas91, Fyo91, Wil91}.
BRMs constitute a synthesis of two natural generalizations of the 1D Anderson
model~(\ref{H1}): on the one hand, they can be used to describe 1D models with
hopping processes linking each site with its first $b$ neighbors; on the other
hand, they can be mapped onto quasi-1D models~\cite{Fyo91}.

Because of the importance of BRMs in the physics of quantum chaos and disordered
systems, it appears natural to ask whether the boomerang effect survives when the
hopping amplitudes are random variables. 
We would like to stress that  it is difficult to predict \textit{a priori} whether the
modification of the quantum dynamics entailed by the hopping processes will
preserve or hinder the boomerang effect.
On the one hand, BRMs represent ``local'' Hamiltonians (remote sites are not
directly connected as is the case for full random matrices), and they share many
features with the standard Anderson model, such as the localization of all
eigenstates (for finite BRMs of size $N \times N$, this is true as long as
$b \ll \sqrt{N}$).
On the other hand, BRMs can be mapped onto quasi-1D models, whose transmission properties are more complex than those
of strictly 1D chains due to the presence of several transmission channels.

To clarify whether the boomerang effect survives in BRM models,
we first considered BRMs of the form
\begin{equation}
     H_{ij} = \delta_{ij} \varepsilon_{i} + (1 - \delta_{ij}) h_{ij},
	\label{brmat}
\end{equation}
where the $\{\varepsilon_{i} \}$ variables have the same uniform
distribution~(\ref{box_dist}) as the site energies in the Anderson
model~(\ref{H1}), while the matrix elements $h_{ij}$ vanish outside
 a band of width $b$,
\begin{equation}
    \begin{array}{ccc}
    	h_{ij} =  0 & \mbox{ if } & |i-j| > b .
    \end{array}
    \label{bandwidth}
\end{equation}
Inside the band, the hopping amplitudes $h_{ij}$ are independent,
identically distributed random variables with box distribution
\begin{equation}
   p(h_{ij}) = \left\{ \begin{array}{cl}
   	1/2W_{b} & \mbox{ if } -W_{b} \leq h_{ij} \leq W_{b} \\
   	0 & \mbox{ otherwise } . \\
   \end{array} \right.
   \label{hop_dist}
\end{equation}
This implies, in particular, that the mean hopping amplitudes are zero,
$\overline{h_{ij}}=0$, a property that will turn out to be crucial in the
following. In our numerical simulations, we set
$\sigma_{\varepsilon}^{2} = W^2/3$ for the random site energies
$\{\varepsilon_{i} \}$, and a weaker disorder 
$W_{b} = 0.1W$ and $\sigma_{b}^{2} = 10^{-2} \sigma_{\varepsilon}^{2}$
for the hopping amplitudes $h_{ij}$. With these parameters, we study the 
temporal evolution of the initial wavepacket~(\ref{init_wp}) with
the Hamiltonian~(\ref{brmat}), for band widths $b=1,2,3$. 

\begin{figure}
\begin{center}
  \includegraphics[width=0.8\linewidth]{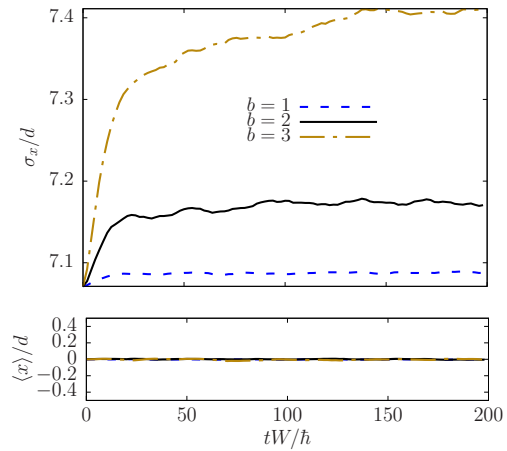}
\caption{\label{fig3}
Evolution of the wavepacket width $\sigma_x$
  (top panel) and
of the center of mass $\langle x(t) \rangle$ (bottom panel) as functions of the
rescaled time $tW/\hbar$, for the BRM model~(\ref{brmat})
for $b=1$, $b=2$, and $b=3$.}
\end{center}
\end{figure}

Our numerical results for the center of mass  of the wavepacket are
displayed in the lower panel of Fig.~\ref{fig3}. They show that, even for
the modest values of $b$ considered, the center of mass does not evolve
in time. Increasing the value of $b$ does not change 
this conclusion.
As will be confirmed below, this behavior is essentially due to the
fact that the random amplitudes $h_{ij}$ with
distribution~(\ref{hop_dist}) have a vanishing average,
$\overline{ h_{ij}} = 0$. This inhibits any drift of the center
of mass and, in particular, prevents the boomerang effect to occur.

The absence of drift, however, does not imply that the quantum particle
is not scattered: in fact, the hopping processes cause the particle to
diffuse around its initial position with a corresponding spread of its
wavefunction. This is demonstrated by the analysis of the second moment
of the density distribution, shown in the upper panel of Fig.~\ref{fig3}.
As in the Anderson model, we find that the wavepacket first spreads
ballistically and then gets localized at long times.
Increasing $b$ produces a larger spread of the wavepacket, as can be expected
considering that the localization length of the eigenvectors of BRMs roughly
scales as $\ell_{loc} \propto b^{2}$~\cite{Cas90b,Wil91,Fyo91}.

To confirm the crucial role played in the boomerang effect by the
deterministic component in the nearest-neighbor hopping amplitudes, we also
considered BRMs with an added ``Laplacian'' term, i.e.,
\begin{equation}
	H_{ij} = \delta_{ij} \varepsilon_{i} + (1 - \delta_{ij}) h_{ij} -
	J \left( \delta_{i,j+1} + \delta_{i,j-1} \right) .
	\label{brmat2}
\end{equation}
In Eq.~(\ref{brmat2}), the site energies $\varepsilon_{i}$ are random variables
with the box distribution~(\ref{box_dist}). As in the previous case, the hopping
amplitudes $h_{ij}$ obey Eq.~(\ref{bandwidth}), i.e., they vanish outside
of a band of width $b$, while within the band they are random variables with
distribution~(\ref{hop_dist}).
For the numerical calculations, we set the variance of site energies to
$\sigma_{\varepsilon}^{2} = J^2/3$ (i.e., $W = J$) and  consider two values of
hopping amplitudes: i) off-diagonal disorder weaker than the diagonal one,
 $\sigma_{b}^{2} = J^2/12$ (i.e., $W_{b} = J/2$) and ii) off-diagonal
and diagonal disorder with the same strength,
$\sigma_{b}^{2} = \sigma_{\varepsilon}^{2} = J^2/3$ (i.e., $W = W_{b} = J$).

\begin{figure}
\begin{center}
\includegraphics[width=0.8\linewidth]{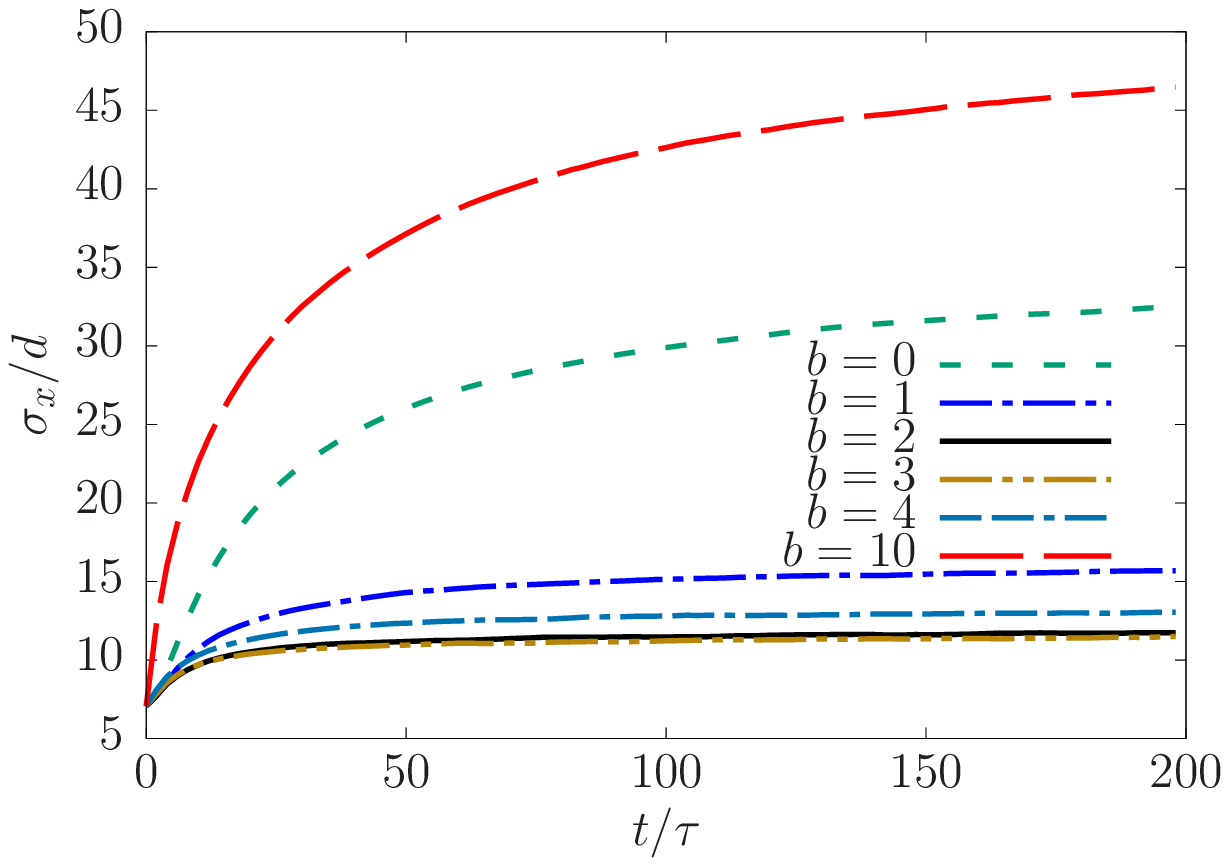}
\includegraphics[width=0.8\linewidth]{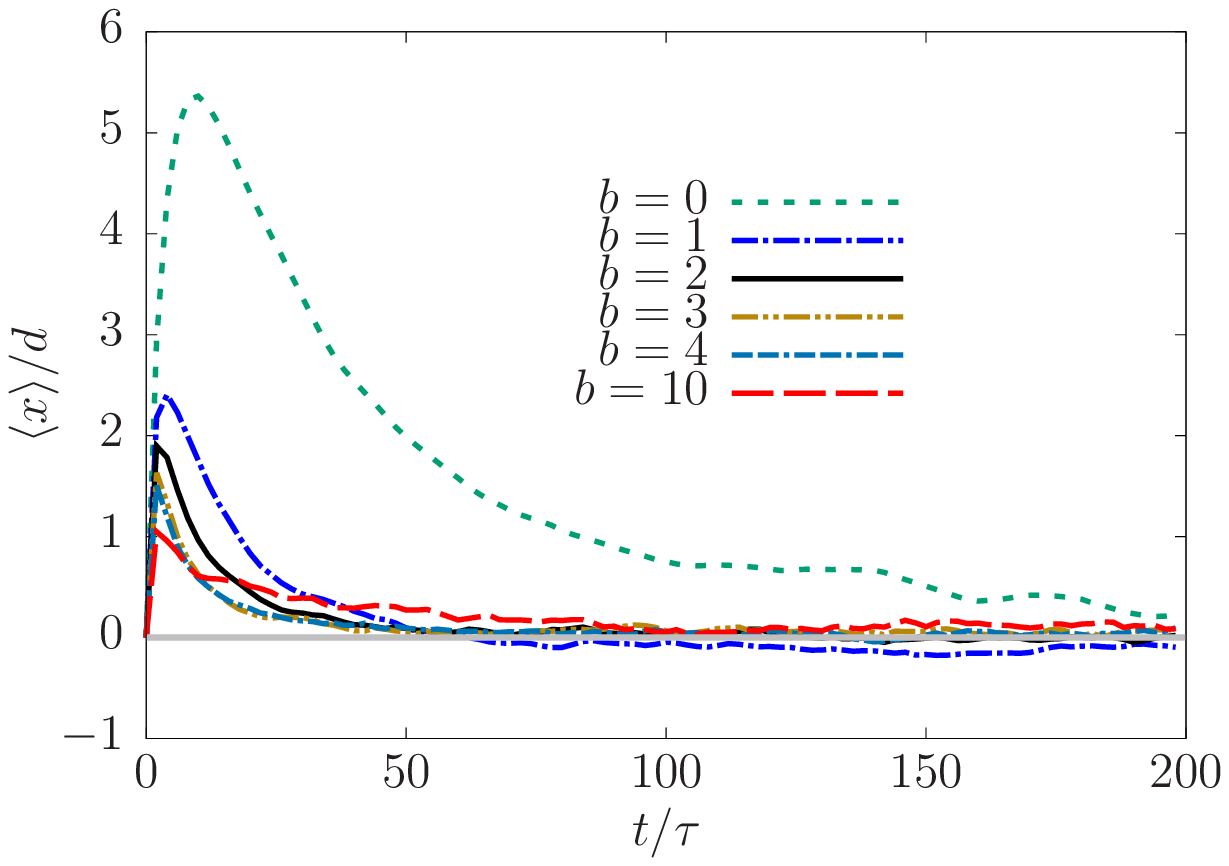}
\caption{\label{fig4}
Wavepacket width $\sigma_x$ (top panel)
  and position of its center of mass $\langle x\rangle$ (bottom panel),
as functions of the rescaled time $t/\tau$, with $\tau=\hbar/J$ for the
BRM model~(\ref{brmat2}). The strength of the diagonal disorder is
$\sigma_{\varepsilon}^{2} = J^2/3$ while the hopping amplitudes have
variance $\sigma_{b}^{2} = J^2/12$. The ensemble average is computed over
$N_{c} = 2000$ disorder realizations.
The grey line in the lower panel marks $\langle x \rangle\!=\!0$.}
\end{center}
\end{figure}

\begin{figure}
\begin{center}
\includegraphics[width=0.8\linewidth]{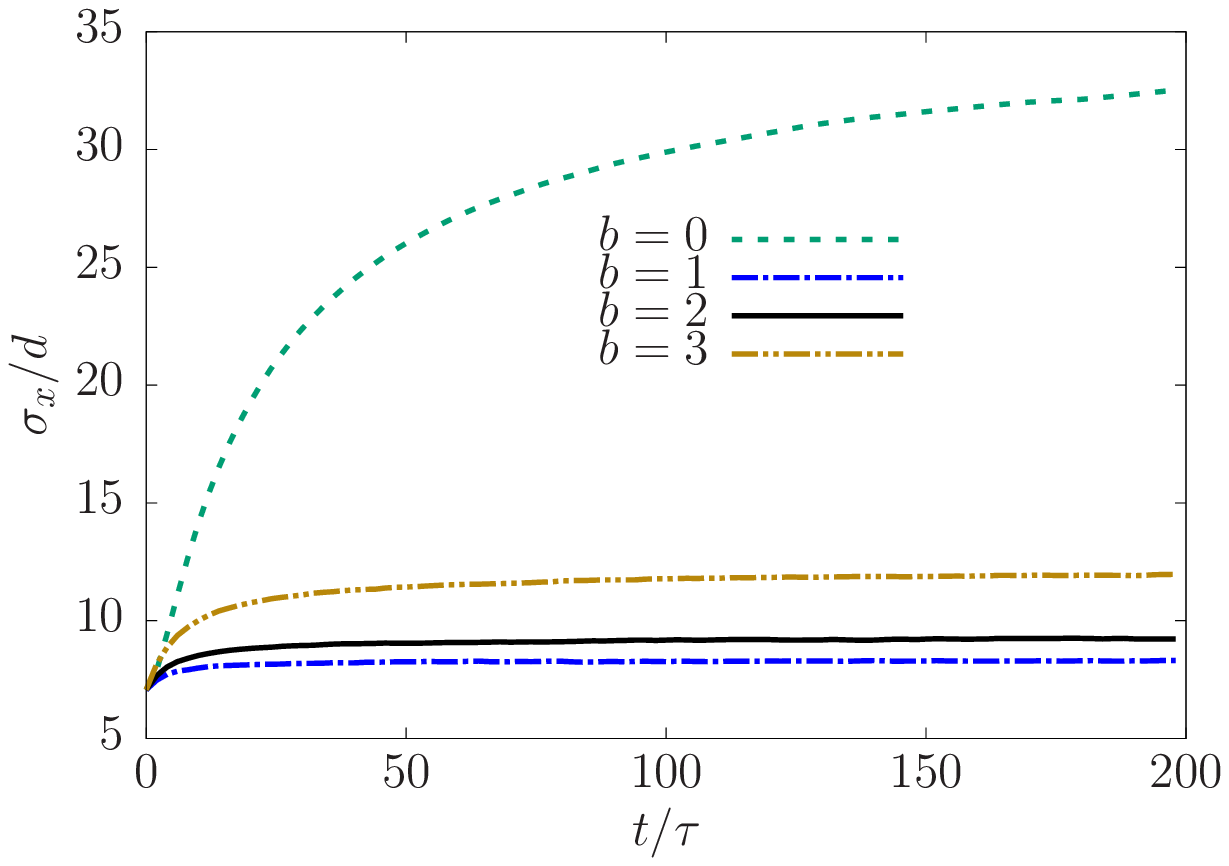}
\includegraphics[width=0.8\linewidth]{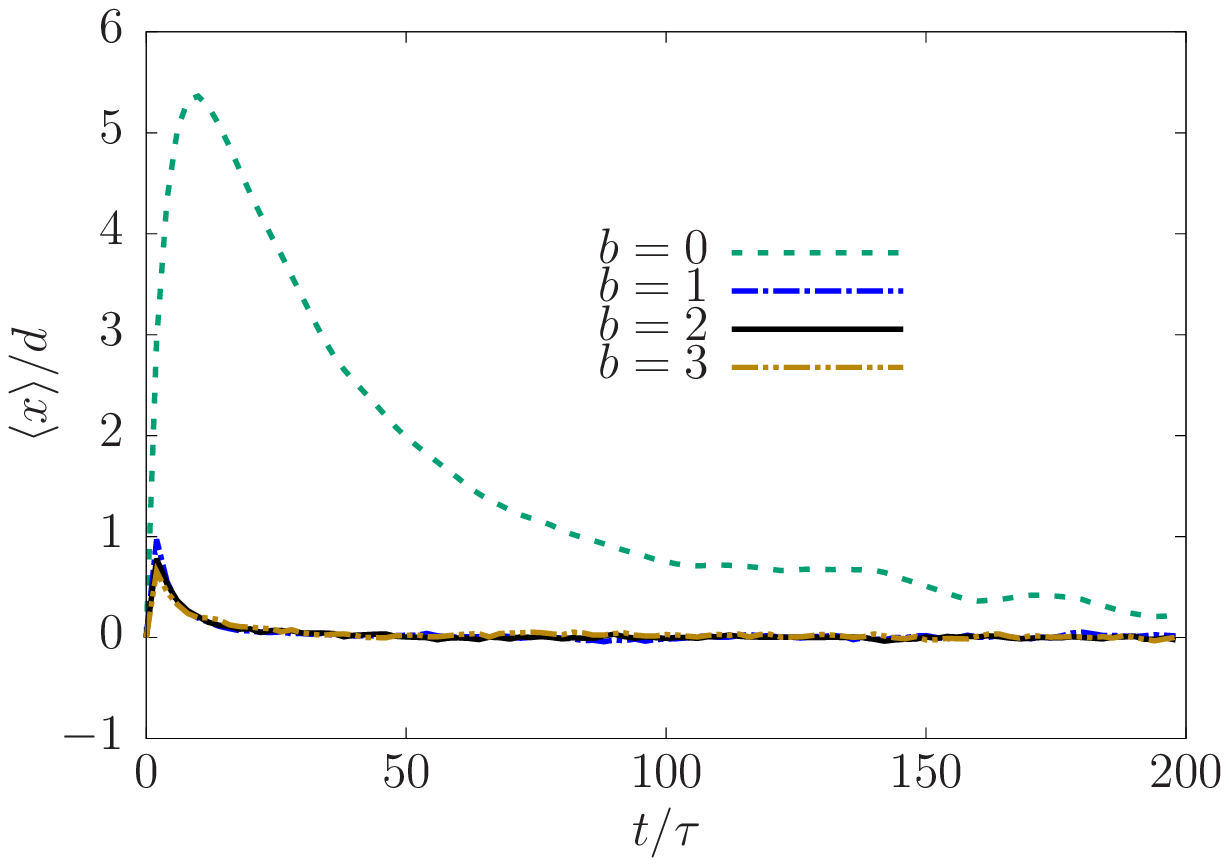}
\caption{\label{fig5}
  Wavepacket width $\sigma_x$ (top panel) and 
  center of mass position $\langle x\rangle$ of the
wavepacket (bottom panel) as functions of the rescaled time $t/\tau$, with
$\tau = \hbar/J$ for the BRM model~(\ref{brmat2}).
Here the strength of the diagonal and off-diagonal disorder are the same,
$\sigma_{\varepsilon}^{2} =\sigma_{b}^{2} = J^2/3$.
The ensemble average is computed over $N_{c} = 2000$ disorder realizations.}
\end{center}
\end{figure}

Our numerical results for the model~(\ref{brmat2}) are displayed in
Figs.~\ref{fig4} and~\ref{fig5}. They show that the Laplacian term
restores the boomerang effect when it is dominant with respect to the
random hopping amplitudes.
Increasing the width of the band nevertheless diminishes the distance covered by the
wavepacket before coming back to its original position and therefore
reduces the boomerang effect, as demonstrated in the lower panel
of Fig.~\ref{fig4}.
Fig.~\ref{fig5} shows what happens when the off-diagonal disorder is stronger: it
quickly dominates over the Laplacian even for $b \sim 1$, and therefore effectively
suppresses the boomerang effect. 

\begin{figure}
\begin{center}
\includegraphics[width=0.95\linewidth]{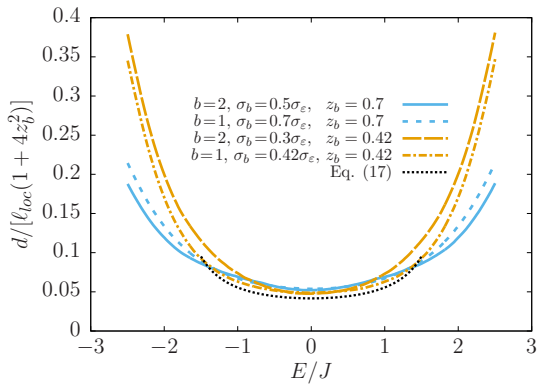}
\caption{\label{figloc}
   Inverse of the rescaled localization length $(\ell_{loc}(1+4z_b^2)^{-1}$ [Eq. (\ref{grosso})]
   in units of $d^{-1}$
   as a function of the energy.
   The numerical data are compared with the rescaled localization length, $\sigma_\varepsilon^2/(8J^2\sin^2(kd))$, as obtained from Eq. (\ref{miaborn}).
Here we set $\sigma_\varepsilon^2=J^2/3$.}
\end{center}
\end{figure}

The study of the second moment of the wavepacket shows that for
BRMs of the form~(\ref{brmat2}), the spreading of the wavefunction
in the localized regime does not increase continuously with $b$, as
one might naively expect. When the band width lies in the range
$b \sim 1-3$, increasing $b$ actually \emph{reduces} the asymptotic
value of $\langle x^{2}(t) \rangle$. For $b > 3$, however,
numerical data suggest that the spatial extension of the wavepacket 
in the localized regime grows with $b$.
To understand the behavior of $\langle x^2(t) \rangle$ for small
values of $b$, we numerically computed the inverse localization
length for the model~(\ref{brmat2}), using the identity~\cite{Farchioni1992,Larcher2013}
\begin{equation}
  \ell_{loc}^{-1} = \lim_{N \to \infty}
  \dfrac{1}{N d} \ln \left|\dfrac{G_{N N}(E)}{G_{1 N}(E)} \right|
  \label{grosso}
\end{equation}  
where
\begin{displaymath}
   G(E) = \frac{1}{E-H}	
\end{displaymath}
is the Green's function of the Hamiltonian~(\ref{brmat2})
with matrix elements $G_{ij}(E) = \langle i|G(E)|j \rangle$.
Using formula~(\ref{grosso}), we computed the inverse localization
length for $b = 1$ and $b = 2$ for various strengths of the
off-diagonal disorder. We set $N= 200$ and we averaged the result
over an ensemble of $N_{c} = 1000$ disorder configurations.
The numerical data suggest that, as long as $b$ is small and the
off-diagonal disorder is weak, at fixed diagonal
disorder strength $\sigma_\varepsilon$, the relative strength of the
off-diagonal disorder with respect to the Laplacian term is
given by the parameter
\begin{displaymath}
	z_b = \sqrt{b} \sigma_{b}/\sigma_\varepsilon .
\end{displaymath}
This is corroborated by the data in Fig.~\ref{figloc}, which
show the behavior of the inverse localization length as a function
of energy for four different values of $b$ and $\sigma_{b}$,
and by their comparison with the analytical expression of the
localization length obtained in the Born
approximation~\cite{Mueller2011} when diagonal and
off-diagonal disorders are
uncorrelated (see Appendix \ref{miaappen}):
\begin{equation}
  \ell_{loc}^{-1}=\dfrac{1}{d}
  \dfrac{\sigma_\varepsilon^2+4b\sigma^2_b}{8J^2\sin^2(kd)}.
  \label{miaborn}
\end{equation}  
We observe that the inverse localization lengths, after being rescaled
by a factor $(1+4z_b^2)$, nearly coincide for the cases $z_b=0.7$ and
$z_b=0.42$ (and for both $b=1$ and 2)
and are in good agreement, at the band center,
with Eq. (\ref{miaborn}).
Specifically, the numerical data and Eq. (\ref{miaborn})
show that the localization length scales with
$1/(1+4z_{b}^2)$
as long as $z_{b} \lesssim 1$. The localization length thus 
decreases with $b$, which agrees with the reduction of the
asymptotic width of the wavepacket for $b \leq 3$ that can be seen from
the top panel of Fig.~\ref{fig4}.
When $z_{b} > 1$ on the other hand, the off-diagonal hopping terms
start to dominate over the Laplacian and $\ell_{loc}$ starts to
increase with $b$
in agreement with the usual behavior of BRMs of the
form~(\ref{brmat}).

\section{Quantum kicked rotor}
\label{kicked}

The kicked rotor is a physical system that has played a key role in the study
of classical and quantum chaos~\cite{Cas79, Izr90, Fishman1982, Casati1989}.
Its realization in cold atom experiments has provided additional reasons of
interest~\cite{Raizen1994, Moo95, Amm98, Rin00, Arc01, Chabe2008}.
The kicked rotor is also closely related to the Anderson model~(\ref{H1}).
From a mathematical point of view, the correspondence between the Anderson model
and the kicked rotor lies in the fact that the Hamiltonian of the former is a
tridiagonal matrix with diagonal disorder, while the latter can be mapped onto a
tight-binding model with pseudo-random diagonal elements~\cite{Fishman1982, She86}.
From a physical perspective, the counterpart of the localization of the eigenstates in the Anderson
model is a suppression of the energy growth in the
kicked rotor, a phenomenon known as ``dynamical localization''.
The close analogy between the kicked rotor and the Anderson model suggests that the
quantum boomerang effect, which exists in the first system, ought to be present also in
the second one. In this section we show that this is indeed the case, although the
boomerang dynamics in the kicked rotor exhibits a specific dependence on the initial
state which has no counterpart in the Anderson model.

The quantum kicked rotor is defined by the Hamiltonian
\begin{equation}
  H = \dfrac{p^{2}}{2} + V(x) \sum_{n=-\infty}^{\infty} \delta \left( t - n \right) ,
  \label{kr_ham}
\end{equation}
with
\begin{displaymath}
	V(x) = K \cos (x) .
\end{displaymath}
It describes a planar rotor periodically subjected to instantaneous variations
of the momentum (``kicks'')  with a period $T = 1$. The parameter $K$ determines the strength of the kicks.

The variable $x$ in the Hamiltonian~(\ref{kr_ham}) can be interpreted either as
an angle or as a spatial Cartesian coordinate. In the first case, one has
$x \in [-\pi,\pi]$ and $p$ is the associated angular momentum. In the second case
$x \in \mathbb{R}$ and $p$ is the ordinary momentum conjugated to a spatial coordinate.
The first interpretation is usually used in the study of classical and quantum
chaos. The second one is more appropriate for the analysis of experiments with cold
atoms in optical lattices (and, for this reason, we refer to the model~(\ref{kr_ham})
with $x \in \mathbb{R}$ as the ``atomic'' kicked rotor).

The correspondence between the kicked rotor~(\ref{kr_ham}) and the Anderson
model~(\ref{H1}) was first established in~\cite{Fishman1982} (see also~\cite{Haa10}). 
Below we recall the main steps of this approach, considering the $x$
variable as an angle for the sake of simplicity. In this case, $p$ is an angular momentum and its
eigenvalues are integer multiples of $\hbar$.
As a first step, it is useful to consider the Floquet operator of the kicked rotor
in the momentum representation:
\begin{equation}
	U(\alpha) = e^{-i p^{2} (1 - \alpha)/2\hbar} e^{-iV/\hbar} 
	e^{-i p^{2} \alpha/2\hbar} .
	\label{floquet}
\end{equation}
The propagator~(\ref{floquet}) describes the evolution over the period
$[n - \alpha, n + 1 -\alpha]$, with $\alpha \in [0,1]$. It is the product of three
terms: the first factor on the right describes the free evolution over the time interval
$[n - \alpha,n]$, the central term represents the kick at $t = n$, while the leftmost
factor gives the free motion over the interval $[n, n + 1 - \alpha]$.
The parameter $\alpha$ defines the time elapsed before the rotor is initially kicked: the
kick occurs at the beginning of the interval if $\alpha = 0$, at the end if
$\alpha = 1$ and in the middle of the period if $\alpha = 1/2$.
If one introduces a new operator $M$ via the equation
\begin{equation}
	e^{-iV/\hbar} = \frac{1 + i M}{1 - i M} ,
	\label{w_op}
\end{equation}
the Floquet operator~(\ref{floquet}) becomes
\begin{equation}
	U(\alpha) = e^{-i p^{2} (1 - \alpha)/2\hbar} \frac{1 + i M}
	{1 - i M} e^{-i p^{2} \alpha/2\hbar} .
	\label{floquet2}
\end{equation}

Let $|\phi_{\alpha} \rangle$ be a Floquet (quasi)-eigenstate, satisfying the equation
\begin{equation}
	U(\alpha) |\phi_{\alpha} \rangle = e^{-i E_{\alpha}/\hbar} |\phi_{\alpha} \rangle .
	\label{flo_eig}
\end{equation}
Using the representation~(\ref{floquet2}) for the Floquet operator, one can write
the previous equation as
\begin{equation}
	e^{-i p^{2} (1 - \alpha)/2\hbar} \frac{1 + i M}{1 - i M}
	e^{-i p^{2} \alpha/2\hbar} |\phi_{\alpha} \rangle = e^{-iE_{\alpha}/\hbar} |\phi_{\alpha}\rangle .
	\label{eigvec}
\end{equation}
After introducing the vector
\begin{displaymath}
	|\psi_{\alpha} \rangle = \frac{1}{1 - i M} e^{-i p^{2} \alpha/2\hbar}|\phi_{\alpha}\rangle ,
\end{displaymath}
one can cast Eq.~(\ref{eigvec}) in the form
\begin{equation}
    e^{-i(p^{2}/2 - E_{\alpha})/2\hbar} (1+ i M) |\psi_{\alpha} \rangle = 
    e^{i(p^{2}/2 - E_{\alpha})/2\hbar} (1 - i M) |\psi_{\alpha} \rangle .
	\label{eigvec2}
\end{equation}

Let $\{ |m \rangle \}$ represent a complete set of eigenstates of the momentum $p$.
If the vector $|\psi_{\alpha} \rangle$ is expanded in the momentum basis, one can write
\begin{equation}
	|\psi_{\alpha} \rangle = \sum_{m} \psi_{m}^{(\alpha)} |m \rangle
	\label{mom_expan}
\end{equation}
with $\psi_{m}^{(\alpha)} = \langle m | \psi_{\alpha} \rangle$.
Substitution of the expansion~(\ref{mom_expan}) in Eq.~(\ref{eigvec2}) gives
\begin{displaymath}
\begin{array}{l}
    \displaystyle
	\sum_{m} e^{-i(p^{2}/2 - E_{\alpha})/2\hbar} (1 + i M)|m\rangle \psi_{m}^{(\alpha)} = \\
	\displaystyle
	\sum_{m} e^{i(p^{2}/2 - E_{\alpha})/2\hbar} (1 - i M)|m\rangle \psi_{m}^{(\alpha)} . \\
\end{array}
\end{displaymath}
Projecting both sides of this equation on the momentum bra $\langle n|$ and rearranging the
terms, one finally obtains
\begin{equation}
	\epsilon_{n} \psi_{n}^{(\alpha)} +
	\sum_{m \neq n} \langle n | M | m \rangle \psi_{m}^{(\alpha)} = E_{0} \psi_{n}^{(\alpha)}
	\label{kr_and}
\end{equation}
where the symbol $\epsilon_{n}$ represents the ``site energies''
\begin{equation}
	\epsilon_{n} = \tan \left[ \frac{1}{2\hbar} \left( E_{\alpha} - \frac{\hbar^{2} n^{2}}{2} \right) \right]
	\label{eps_var}
\end{equation}
while the zero-th component of the $M$ operator plays the role of the energy $E_{0} =
- \langle 0| M | 0 \rangle$.

Eq.~(\ref{kr_and}) shows that the kicked rotor~(\ref{kr_ham}) can
be mapped onto a tight-binding model whose Hamiltonian is an effective
band matrix with pseudo-random diagonal disorder.
Indeed, the variables $\{\epsilon_{n}\}$ represent the site energies and are pseudo-random variables
with Lorentzian distribution, while the terms $\langle n|M|m \rangle$ provide the hopping
amplitudes.
The matrix elements $\langle n | M | m \rangle$ can be calculated in closed
form for $K/\hbar < \pi$ and they fall off exponentially for increasing values
of $| n - m |$.
The above mapping suggests that the kicked rotor might behave as the
BRM models considered in Sec.~\ref{bands}. However, two differences exist between the
two models of the previous section and the tight-binding model~(\ref{kr_and}): the site
energies~(\ref{eps_var}) are not truly random variables and, in
addition, the hopping terms $\langle n|M|m \rangle$ are deterministic constants.
From this point of view, the tight-binding model~(\ref{kr_and}) is closer
to the pseudo-random Anderson model considered in Sec.~\ref{pseudo}; one can therefore
expect that the boomerang effect should occur in the kicked rotor model~(\ref{kr_ham}).

To check whether this conclusion is correct, we numerically evaluate the evolution of
a Gaussian wavepacket (in momentum space) with Hamiltonian~(\ref{kr_ham}), with the variable
$x$ spanning the real axis. In this case, the spatial potential in the Hamiltonian~(\ref{kr_ham})
is $(2 \pi)$-periodic and the Bloch theorem applies. As a consequence, the eigenstates of the
momentum $p$ are now defined by an integer quantum number $n$ and a real quasi-momentum
$\beta$ in the interval [-1/2;1/2(. In other words, one has
\begin{displaymath}
	p |n, \beta \rangle = \hbar (n + \beta) |n, \beta \rangle .
\end{displaymath}
Note that, since the dynamical localization of the kicked rotor occurs in momentum
space, the analysis of Sec.~\ref{themodel} must now be transposed from the $x$- to
the $p$-space. For this purpose, we consider an initial wavepacket of the form
\begin{equation}
    \begin{array}{ccl}
	\psi_{n,\beta}(t=0) & = & \langle n, \beta| \psi(t=0) \rangle \\
	& = & \displaystyle {\cal{N}} \exp \left[-\frac{\hbar^2(n + \beta)^{2}}{2 \sigma_{p}^{2}(0)}
	- i (n + \beta) x_{0} \right] , \\
	\end{array}
	\label{init_wp2}
\end{equation}
where $\cal{N}$ is a normalization constant (approximatively equal to
${\cal{N}} \simeq \hbar/\sqrt{\pi \sigma_{p}^{2}(0)}$ if $\sigma_{p}(0) \gg 1$), while
the parameter $\sigma_{p}(0)/\sqrt{2}$ gives the width of the wavepacket in momentum space,
which we chose much larger than $\hbar$.
This implies that the wavefunction in the coordinate representation is a narrow Gaussian
with variance $\sigma_{x}^{2}(0) = \hbar^{2}/2\sigma_{p}^{2}(0)$. The parameter $x_{0}$
represents the initial ``boost'' of the packet.
To numerically propagate the initial state~(\ref{init_wp2}),
we used the quantum map
\begin{equation}
	|\psi(t+1) \rangle = U(\alpha) |\psi(t) \rangle
	\label{q_map}
\end{equation}
where $U(\alpha)$ is the Floquet operator~(\ref{floquet}).
In the momentum representation, its matrix elements take the form
\begin{equation}
    \begin{array}{l}
	\langle n, \beta | U(\alpha) | m, \beta^{\prime} \rangle  
	=
	i^{m-n} e^{-i \hbar (n + \beta)^{2}(1 - \alpha)/2}\\
	\times J_{n-m}(K/\hbar) e^{-i \hbar (m+\beta^{\prime})^{2} \alpha/2}
	\delta(\beta - \beta^{\prime}) ,
	\end{array}
	\label{evol_mat}
\end{equation}
where $J_{n}(k)$ is a Bessel function of the first kind, with integral representation
\begin{displaymath}
	J_{n}(k) = \frac{1}{\pi i^{n}} \int_{0}^{\pi} \mathrm{d} \theta e^{i k \cos \theta}
	\cos n \theta .
\end{displaymath}
Notice that the Bessel functions decrease quickly when the index becomes larger than the
argument; this entails that the elements of $U$ fall off for $|n~-~m|~\gtrsim~K/\hbar$
and that the matrix~(\ref{evol_mat}) has an effective band structure. The phase factors, on
the other hand, endow the matrix $U_{nm}$ with a pseudo-random character.

\begin{figure}
  \begin{center}
\includegraphics[width=0.95\linewidth]{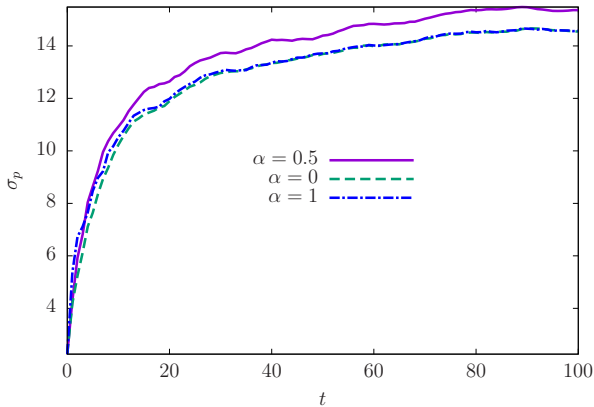}
\caption{\label{fig-width}Width of the momentum distribution 
  $\sigma_p$
  as a function of time, computed numerically for the kicked rotor starting from the initial state (\ref{init_wp2}) and with $\hbar\!=\!1$. The violet continuous line corresponds to
  $\alpha=0.5$, the blue dot-dashed line to $\alpha=1$, and
  the green dashed line to $\alpha=0$.
  Here we set $x_0=\pi/2$
  and $\sigma_{p}(0)=10$. The average is done over 1000
values of $\beta$.}
\end{center}
\end{figure}

\begin{figure}[h]
\begin{center}
\includegraphics[width=0.95\linewidth]{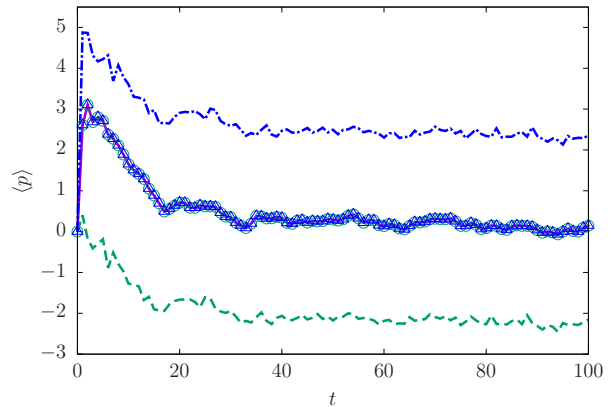}
\caption{\label{fig-kick}Average momentum $\langle p\rangle$
  as a function of time, computed numerically for the kicked rotor starting from the initial state (\ref{init_wp2}) and with $\hbar\!=\!1$.  Here we set
  $x_0=\pi/2$
  and $\sigma_{p}(0)=10$. The violet continuous line corresponds to
  $\alpha=0.5$, the blue dot-dashed line to $\alpha=1$, and
  the green dashed line to $\alpha=0$. The green circles and the blue triangles
  correspond to $\alpha=0$ and $\alpha=1$ respectively,
  for the case of an initial state dephased according to Eq. (\ref{q_state}). The averages are performed over 1000
values of $\beta$.}
\end{center}
\end{figure}

In our numerical computations, we took $\hbar\!=\!1$ and we considered the initial state~(\ref{init_wp2}) with
$x_{0} = \pi/2$ and $\sigma_{p}(0) = 10$.
Following~\cite{Lemarie2017}, we averaged the time evolution of the initial wavepacket
over $N_{\beta} = 1000$ values of the quasi-momentum $\beta$. 
In the results shown below, we set the strength $K$ of the kicking potential to $K = 5$, which corresponds
to the region of strong chaos for the classical kicked rotor. 
We selected three values for the parameter $\alpha$: $\alpha = 0$ (kick followed by free
evolution over a period), $\alpha = 1$ (free evolution over a period and then a kick),
and $\alpha = 1/2$ (kick preceded and followed by half a period of free evolution).
Figs.~\ref{fig-width} and~\ref{fig-kick} show the results obtained for the wavepacket width $\sigma_p(t)$ (which measures the kinetic energy of the kicked rotor) and the mean wavepacket position $\langle p(t) \rangle$ in momentum space.
Fig.~\ref{fig-width} shows that the kinetic energy of the system first increases quickly but then slows down. This corresponds to  localization in momentum
space of the wavepacket~(\ref{init_wp2}) and is known as dynamical
localization. Varying the parameter $\alpha$ does not produce significant
differences in the behavior of the energy, except for a small increase of its long-time value for $\alpha=1/2$.
The situation is quite different for the temporal evolution
of $\langle p(t) \rangle$.
As can be seen from Fig.~\ref{fig-kick}, when $\alpha = 1/2$ a quantum boomerang effect
is present: the center of mass of the wavepacket first moves away from the
origin, and eventually comes back to its initial position.
However, when $\alpha \neq 1/2$, the center of the wavepacket does not
return to the starting point but instead gets localized in a different position,
to the left (for $0 \leq \alpha < 1/2$) or to the right of the origin (for
$1/2 < \alpha \leq 1$). The asymptotic value of $\langle p(t) \rangle$ increases
continuously with $\alpha$. We have also performed simulations for different values of $K$ ranging in the interval  $1-10$ (not shown) and have found qualitatively similar results.

Two remarks are in order concerning the dependence on $\alpha$ of the the long-time value of $\langle p(t) \rangle$.
First, we observe that selecting $\alpha = 1/2$ endows
the quantum map~(\ref{q_map}) with the symmetry under time reversal
that is required for the boomerang effect to appear~\cite{Prat2019}.
Indeed, the time evolution described by $U(1/2)$ consists of a kick preceded
and followed by an half-period of free evolution, so that moving forwards or
backwards in time is completely equivalent.
This is no longer true for every other value of $\alpha$: for instance,
if $\alpha = 0$ the evolution towards positive times starts with a kick followed
by free motion, whereas the evolution towards negative times has the free motion
preceding the kick. This is the reason why the center of the wavepacket does
not come back to its original position when $\alpha \neq 1/2$.

As a second remark, we observe that the shift of the asymptotic position
of the wavepacket is due to the rightmost factor in Eq.~(\ref{floquet}).
This can be seen as follows. It is easy to show that two Floquet operators,
corresponding to different values of $\alpha$, are related by the
identity
\begin{displaymath}
	U(\alpha_{2}) = e^{-ip^{2} \Delta \alpha/2\hbar} U(\alpha_{1})
	e^{ip^{2} \Delta \alpha/2\hbar}
\end{displaymath}
with $\Delta \alpha = \alpha_{2} - \alpha_{1}$.
The same relation holds for their $N$-th powers
\begin{equation}
	\left[ U(\alpha_{2}) \right]^{N} = e^{-ip^{2} \Delta \alpha/2\hbar}
	\left[ U(\alpha_{1}) \right]^{N} e^{ip^{2} \Delta \alpha/2\hbar} .
	\label{u2}
\end{equation}
Applying both sides of Eq.~(\ref{u2}) to an initial state $|\Psi(0)\rangle$,
and projecting the resulting vectors onto the $|m,\beta \rangle$ momentum
eigenstate, one obtains
\begin{equation}
  \left| \langle m, \beta | \left[ U(\alpha_{2}) \right]^{N} | \Psi(0) \rangle \right|^{2} \!=\!
	\left| \langle m, \beta | \left[ U(\alpha_{1}) \right]^{N} | \Phi(0) \rangle \right|^{2}
	\label{prob}
\end{equation}
with
\begin{equation}
	|\Phi(0) \rangle = e^{i p^{2} \Delta \alpha/2\hbar} | \Psi(0) \rangle .
	\label{q_state}
\end{equation}
Eq.~(\ref{prob}) shows that letting an initial state $|\Psi(0)\rangle$ evolve with
the quantum map~(\ref{q_map}) with $\alpha = \alpha_{2}$ produces a quantum state
with the same probability distribution as the state obtained by first applying the
operator $e^{i p^{2} \Delta \alpha/2\hbar}$ to the initial state $|\Psi(0)\rangle$ and
then letting it evolve with the quantum map~(\ref{q_map}) with $\alpha = \alpha_{1}$.
This conclusion is confirmed by numerical calculations shown in Fig.~\ref{fig-kick}:
by using Eq.~(\ref{q_state}) to dephase the momentum components of the initial
wavepacket~(\ref{init_wp2}) and letting the resulting state evolve with the
map~(\ref{q_map}) with $\alpha = 0$ and $\alpha = 1$, the boomerang dynamics becomes
identical to that observed for $\alpha=1/2$ without dephasing. 
This shows that the dynamical evolution of the center of mass in the kicked rotor can be
controlled by appropriately tayloring the initial state. This also agrees with previous
observations of the dependence of the dynamics of the kicked
rotor on the initial state~\cite{Bru01a,Bru01b,Sadgrove2007,Dana2008,Delvecchio2020}.

To conclude our study of the kicked rotor, we also investigated the relevance of the
pseudo-random character of the phase factors in the evolution matrix~(\ref{evol_mat}).
To this end, we replaced $\hbar n^{2}/2$ and $\hbar m^{2}/2$ in Eq.~(\ref{evol_mat}) with uncorrelated
random phases $\phi_{n}$ and $\phi_{m}$, uniformly distributed in the interval $[0,2\pi[$.
The system thus obtained constitutes a purely random kicked rotor. We found that the evolution of
$\langle p(t) \rangle$ has the same behavior observed in the kicked rotor. In particular,
the initial state~(\ref{init_wp2}) has a boomerang dynamics only if $\alpha = 1/2$.
For different values of $\alpha$, the asymptotic value of $\langle p(t) \rangle$ again does not
vanish, unless the initial state is modified with an appropriate change of the
phases of the momentum components.

\section{Conclusions}
\label{concl}

The purpose of this work was to assess the robustness of the quantum boomerang effect
in various random and pseudo-random tight-binding models commonly used
in the theory of low-dimensional disordered systems. We also considered
a closely related model, i.e., the kicked rotor, which has played a crucial
role in the study of quantum chaos.

Our findings show that the quantum boomerang effect is a rather widespread phenomenon that can be found in every tight-binding
model with diagonal disorder and deterministic hopping amplitudes. The random
or pseudo-random character of the site energies does not seem to make a big
difference.
On the other hand, the introduction of hopping processes with zero-average
random amplitudes suppresses the boomerang dynamics.
Our study of the kicked rotor, finally, shows that the boomerang effect can be
observed also in this model, although with a specific dependence on the initial
state which has no analog in the Anderson model.
We can therefore conclude that the boomerang effect is not a specificity of the
Anderson model, but a general feature that can be observed in a broad variety of tight-binding
models with diagonal disorder. Interesting open questions include the fate of this phenomenon in other symmetry classes -- for example when time reversal invariance is broken -- or in interacting systems \cite{Janarek20}.

\section*{Acknowledgements}
P.V. acknowledges the Laboratoire Kastler Brossel and the Piri Reis University
for their hospitality.  N.C. and D.D acknowledge financial support from the Agence Nationale
de la Recherche (grants ANR-19-CE30-0028-01 CONFOCAL and ANR-18-CE30-0017 MANYLOK, respectively).
L.T. acknowledges the financial support of the UMSNH-CIC 2021 grant.

\appendix
\section{Derivation of Eq. (\ref{miaborn})}
\label{miaappen}
In this appendix we provide a derivation of the expression~(\ref{miaborn}) for the
inverse localization length of the eigenstates of the model~(\ref{brmat2}).
For a weak short-range disorder in a 1D system,
the localization length is directly related to the mean free path through
the equation \cite{Beenakker1997,Mueller2011}

\begin{equation}
  \dfrac{1}{\ell_{loc}}=\dfrac{1}{4\ell}=\dfrac{1}{4v\tau}
  \label{primopunto}
\end{equation}
with
\begin{equation}
  \dfrac{1}{\tau}=-\dfrac{2}{\hbar}{\rm Im}[\mathcal{E}(k)] .
  \label{tauu}
\end{equation}
Remark that in our system, because we take all disorder matrix elements as delta-correlated, the transport mean free path is equal to the scattering one.

In Eq. (\ref{tauu}),
$\mathcal{E}(k)$ represents the self-energy which, in the Born
approximation, can  be written as
\begin{equation}
   \mathcal{E}(k)=\langle k|\overline{VG_0V}|k\rangle ,
   \label{self_energy}
\end{equation}
where $G_{0} = (E-H_0)^{-1}$ is the Green function corresponding to the
unperturbed Hamiltonian $H_0$ defined by Eq. (\ref{H1}), while $V = H - H_{0}$ represents the
difference between the Hamiltonians~(\ref{brmat2}) and~(\ref{H1}). 
By expanding the self-energy~(\ref{self_energy}) on the site basis, one obtains
\begin{equation}
\mathcal{E}(k)=\sum_{j,l,m,n}e^{ik(n-j)}\overline{V_{j,l}[G_0]_{l,m}V_{m,n}}.
\end{equation}
Taking into account that $[G_0]_{l,m}=[G_0]_{l,l}e^{-ik|m-l|}=[G_0]_{0,0}e^{-ik|m-l|}$
and that the non-vanishing averages $\overline{V_{j,l}V_{m,n}}$ are: $\overline{V_{j,j+r}^2}$, $\overline{V_{j+r,j}^2}$, $\overline{V_{j,j+r}V_{j+r,j}}$,
with $r=-b,\ldots,0,\ldots,b$, one gets
\begin{equation}
\mathcal{E}(k)=[G_0]_{0,0}\left[ \overline{V_{j,j}^2}+\sum_{r=-b}^b (\overline{V_{j,j+r}^2}+\overline{V_{j,j+r}V_{j+r,j}})\right].
\label{midi}
\end{equation}
Since $\overline{V_{j,j}^2}=\sigma_\varepsilon^2$ and
$\overline{V_{j,j+r}^2}=\overline{V_{j,j+r}V_{j+r,j}}=\sigma_b^2$,
Eq. (\ref{midi}) can be written
\begin{equation}
  \mathcal{E}(k)=[G_0]_{0,0}(\sigma_\varepsilon^2+4b\sigma_b^2).
  \label{midi2}
  \end{equation}
Finally, putting together Eqs. (\ref{primopunto}), (\ref{tauu}) and (\ref{midi2})
and using the identities
$-2{\rm Im [G_0]_{0,0}}=[J\sin(kd)]^{-1}$ and $v=(2J/\hbar)\sin{kd}$,
one obtains Eq.~(\ref{miaborn}) in a straightforward way.


\end{document}